\pdfoutput=1
\documentclass[11pt]{article}
\usepackage[preprint]{acl}
\usepackage{times}
\usepackage{latexsym}
\usepackage{booktabs}
\usepackage{amsmath}

\usepackage[T1]{fontenc}

\usepackage[utf8]{inputenc}

\usepackage{microtype}

\usepackage{inconsolata}
\usepackage{amssymb}
\usepackage{graphicx}

\title{Generative Retrieval with Preference Optimization for E-commerce Search}

\author{Mingming Li\textsuperscript{\rm 1,\textdagger}
 \and Huimu Wang \textsuperscript{\rm 1,\textdagger} \and  Zuxu Chen\textsuperscript{\rm 2} \and Guangtao Nie\textsuperscript{\rm 1} \\ 
 \bf{Yiming Qiu\textsuperscript{\rm 1}  \and  Guoyu Tang\textsuperscript{\rm 1}  \and  Jingwei Zhuo\textsuperscript{\rm 1,\thanks{Corresponding Author. \textdagger Equal Contribution.}}} \\
 \textsuperscript{\rm 1}JD.com, Beijing, China \\
   \textsuperscript{\rm 2}Shenzhen International Graduate School, Tsinghua University, Beijing, China \\ 
 {chen-zx22@mails.tsinghua.edu.cn} \\
  \{limingming65, wanghuimu1, zhuojingwei1\}@jd.com}
  
\begin{document}
\maketitle

\begin{abstract}
Generative retrieval introduces a groundbreaking paradigm to document retrieval by directly generating the identifier of a pertinent document in response to a specific query. 
This paradigm has demonstrated considerable benefits and potential, particularly in representation and generalization capabilities, within the context of large language models. However, it faces significant challenges in E-commerce search scenarios, including the complexity of generating detailed item titles from brief queries, the presence of noise in item titles with weak language order, issues with long-tail queries, and the interpretability of results.
To address these challenges, we have developed an innovative framework for E-commerce search, called generative retrieval with preference optimization. This framework is designed to effectively learn and align an autoregressive model with target data, subsequently generating the final item through constraint-based beam search. 
By employing multi-span identifiers to represent raw item titles and transforming the task of generating titles from queries into the task of generating multi-span identifiers from queries, we aim to simplify the generation process.
The framework further aligns with human preferences using click data and employs a constrained search method to identify key spans for retrieving the final item, thereby enhancing result interpretability.
Our extensive experiments show that this framework achieves competitive performance on a real-world dataset, and online A/B tests demonstrate the superiority and effectiveness in improving conversion gains. 
 
\end{abstract}

\section{Introduction}

Deep semantic retrieval models~\cite{dpsr,devlin2018bert,qiu2022pre,khattab2020colbert,poeem,li2023adaptive,WangLZZWXLY23,mixqp}, have achieved significant success in online E-commerce retrieval and recommendation systems.
Traditional methods rely on the dual-encoder to learn the dense representations of queries and items. They use the dot-product similarity to measure the relevance between the query and candidate items, but lack fine-grained interactions, leading to sub-optimal performance.
\begin{figure*}[t]
    \centering
    \includegraphics[scale=0.46]{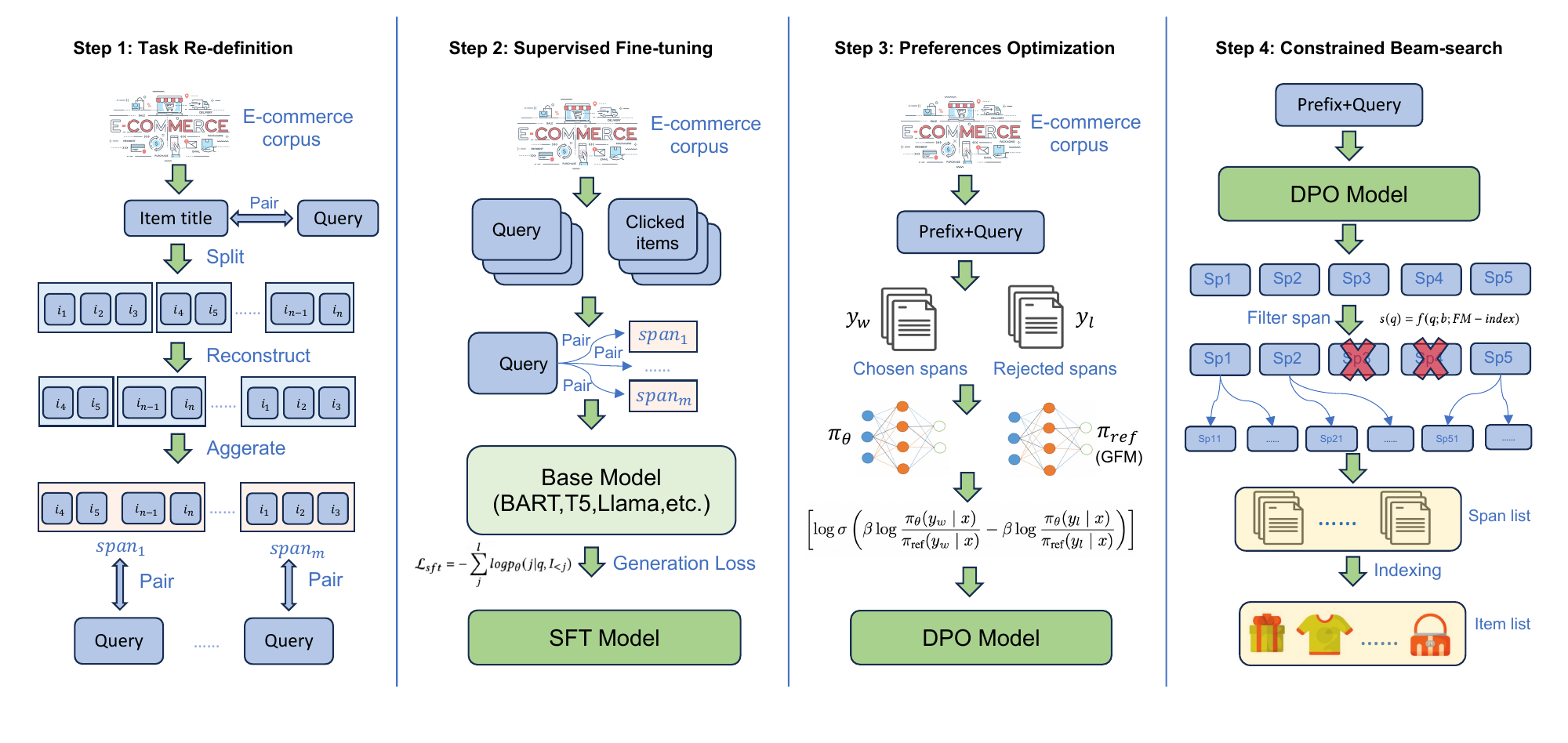}
    \caption{The framework of GenR-PO. It comprises four key stages:
1) Task re-definition stage; 2) Supervised fine-tuning stage; 3) Preference alignment stage; and 4) Inference stage based on constrained beam-search.}
    \label{framework}
\end{figure*}
Recently, a new paradigm,  generative retrieval \cite{NCI,DSI,se-DSI-tang2023semantic,xiaohongshu-semanticid-yuan2024generative,seal-bevilacqua2022autoregressive,rajput2024recommender,ppo-zhou2023enhancing}, has been proposed in the recommendation field and question-answer fields.
These models advocate generating identifiers of target passages/items directly through the autoregressive language models. 
Existing work could be divided into two categories based on identifier types:
1) Numeric-based \cite{NCI,zhuang2022bridging,rajput2024recommender,xiaohongshu-semanticid-yuan2024generative}, they assign numeric identifiers in various ways, e.g., atomic, naive, and semantic. 
2) lexical identifier-based methods \cite{seal-bevilacqua2022autoregressive,glen-lee2023glen,li2023multiview} using the n-grams, title, and URLs as the document identifiers. They could leverage the knowledge of PLMs to decode identifiers, exploring the benefit of pre-trained vocabulary space.
The lexical identifier-based methods show potential in terms of interpretability and generalization capabilities, especially in the era of large language models. Thus, we continue to explore along these lines of methods in this paper.

In the field of E-commerce, there exist several crucial challenges. 
Firstly, the task of query2title generation poses difficulties. Specifically, product titles tend to be lengthy on average, whereas user-entered query words are typically short. Attempting to directly generate lengthy titles can result in significant hallucination issues. While some efforts have been made to utilize pre-trained semantic IDs as document identifiers \cite{NCI,DSI,xiaohongshu-semanticid-yuan2024generative} to simplify the task into query-to-semanticID and reduce complexity, this approach heavily relies on external document representations, deviating significantly from the language itself and necessitating additional calibration, thereby diminishing result interpretability.

Secondly, noise in item titles and weak language order (i.e., keyword stacking) are prevalent issues. In actual product websites, merchants usually provide item titles that contain noise and redundant information. Moreover, the semantic order is predominantly local rather than globally coherent. Essential information such as brand words, attribute words, and categories is often present in the text without regard for position. 

Thirdly, long-tail query challenges are apparent. Unlike in traditional question-and-answer domains, E-commerce faces a severe sample imbalance between queries and items. While some long-tail queries have limited associated products, head queries are linked to a vast array of items. In the age of deep semantics, one-to-many mapping issues can be mitigated through spatial clustering; however, in generative paradigms, such relationships manifest diversely, posing ongoing challenges in resolving them effectively.

Lastly, the interpretability of results is a critical concern. The ability to interpret search results provides valuable insight for enhancing user experience. Unfortunately, deep semantic methods often fall short in this aspect. 

To alleviate the above problems, we introduce a novel framework for E-commerce search, called generative retrieval with preference optimization (GenR-PO).
This framework comprises four key stages: 1) Task re-definition stage; 2) Supervised fine-tuning stage; 3) Preference alignment stage; and 4) Inference stage based on constraint beam-search. More precisely, in the task re-definition stage, we re-construct the item title via word segmentation, sorting, and origination, without losing core information. 
Subsequently, we split the new title into several spans and transformed the query-to-title (query2title) task into a query-to-multi-span (query2multi-span) task, simplifying the generation process due to more sharing spans appearing. 
The supervised fine-tuning stage aims to learn the knowledge of the E-commerce field and reduce illusions. 
The stage of preference optimization is to align with human preference data to produce more significant and human-standard-compliant results. The constraint beam search could prevent the generation of invalid identifiers (i.e., span not occurring in any items). 
 
The contributions of this paper can be summarized as follows:
\begin{itemize}
    \item We discuss the core challenges and redefinition of the generation task to a simple process for E-commerce dense retrieval.

    \item We propose a novel framework, generative retrieval with preference optimization (GenR-PO), that provides a complete pipeline for training, aligning, and inference, meanwhile enhancing result interpretability.
    
    \item We conduct extensive experiments on a real-world dataset. Experimental results demonstrate that GenR-PO achieves significant improvements, especially on long-tail queries, compared to generative retrieval and dense retrieval baselines.

\end{itemize}

\section{Approach}

In this section, we will describe the complete framework, shown in \ref{framework}. It contains four key stages, including task re-definition, Supervised fine-tuning, preference optimization, and constrained beam-search. 

\subsection{Task Re-definition}
Building on the previous discussion, we propose reformulating the raw item's title with and employing a multi-span identifier, which converts the challenging task of matching queries with long titles into a task of associating queries with multiple related text segments. 

Assuming that there is a training sample pair <query, item>, and the item's title consists of $n$ tokens, i.e., $[i_1, i_2, \cdots, i_n]$.  We first adopt the tools of word segment to split the raw title into serval n-grams $\left\{ [i_1,i_2,i_3], [i_4,i_5],\cdots,[i_{n-1}, i_n]\right\}$, including branch name, product name, size word, etc.  Considering the un-sensitivity of position among n-grams, we reconstruct the title via sorting, denoted as 
$\left \{ [i_4,i_5], [i_{n-1},i_n],\cdots, [i_1, i_2, i_3]\right\}$. 
After that, we split the sorted title into serval spans,
\begin{equation}
    \left\{ \underbrace{[i_4,i_5,i_{n-1},i_n]}_{span_1},\cdots, \underbrace{[i_1,i_2,i_3]}_{span_m},\right\}
\end{equation}
where each span has a corresponding length  $l$.   

Following the above reconstruction, we have not only preserved the original information but also augmented the shared information between products at the span level, significantly reducing the noise and diversity introduced by the word order.

Each new span contains a portion of the effective information, representing a perspective. This is because we have simplified the original query2title task into a parallel query2multi-span task, meaning one training sample has become $m$ samples, as follows: 

\[
\begin{array}{c}
   <query, item> \\
  \downarrow \\
  <query, span_1>,\cdots,<query, span_m>
\end{array}
\]

\begin{table*}[tp]
  \caption{Performance of different methods in various types of query. Based on the number of items under each query (\#item), we divide the queries into five categories. The fewer the number, the more long-tailed the description. w/o cons denotes the without constrainted beam-search. The bolded values indicate the optimal values; the underlined values denote the suboptimal values.}
  \label{tab1}
  \centering
  \begin{tabular}{cccccl}
    \toprule
    Method &   \#item =1 &   1<\#item<=5  & 5<\#item<=20 & 20<\#item<=40 & \#item >40 \\
    \midrule
    \multicolumn{6}{c}{Recall@500} \\
    \midrule
    RSR  &  0.2900 & 0.2922 & 0.3083 & \textbf{0.3025} & \textbf{0.2117} \\
     SEAL + SFT  &  0.0180 & 0.0120 & 0.0133 & 0.0102 & 0.0039\\
     TIGER & 0.1470 & 0.1484 & 0.1561 & 0.1801 & 0.1377 \\
    \midrule
    GenR-PO + SFT  & \textbf{0.3760} & \textbf{0.3762} & \underline{0.3266} & 0.2850 & 0.1635\\
    + DPO(w/o cons) &  0.3240 & 0.3344 & 0.3016 & 0.2662 & 0.1544\\
      + DPO (w/ cons)&  \underline{0.3680} & \underline{0.3672} & \textbf{0.3289} & \underline{0.2918} & \underline{0.1690}\\
  \bottomrule
   \multicolumn{6}{c}{Recall@1000} \\
  \midrule
    RSR  &  0.3100 & 0.3086 & 0.3306 & 0.3315 & \textbf{0.2451} \\
      SEAL + SFT &  0.0240 & 0.0198 & 0.0179 & 0.0139 & 0.0061\\
    TIGER  & 0.1920 & 0.1930 & 0.1998 & 0.2307 & 0.1906    \\

    \midrule
     GenR-PO + SFT  &  \underline{0.4230} & \textbf{0.4304} & \underline{0.3890} & 0.3515 & 0.2169\\
     + DPO (w/o cons)&  0.3700 & 0.3803 & 0.3609 & 0.3273 & 0.2074\\
    + DPO (w/ cons)   &  \textbf{0.4310} & \underline{0.4273} & \textbf{0.4001} & \textbf{0.3674} & \underline{0.2330} \\

  \bottomrule
\end{tabular}
\end{table*}

\subsection{Supervised Fine-tuning}
Due to the general pre-trained model lacking e-commerce domain knowledge, we perform supervised fine-tuning (SFT) on specific data via the click pairs between the query and item.
Specifically, for each training sample, the objective is to minimize the sum of the negative log-likelihoods of the tokens $\{i_1,\cdot,i_j,\cdot, i_l\}$ in a target identifier I (span), whose length is $l$. The generation loss is formulated as,
 
\begin{equation}
   \mathcal{L}_{sft} = - \sum_{span}^m \sum_j^{l} \log {p_\theta}(j|q,I_{<j}) 
\end{equation}
where $I_{<j} = \{i_1, i_2,\cdots,i_j\}$,  $p_\theta $ is the SFT model.

\subsection{Preferences Optimization}

Although the supervised fine-tuning model has achieved tremendous success, the outcomes it generates remain uncontrollable, unstable, and do not align with human preference requirements. To alleviate this problem, existing works attempt to align preferences with reinforcement learning from human feedback (RLHF).  However, this pipeline may be too complex and often unstable. Fortunately, recent work DPO \cite{DPO} derives a simple approach for policy optimization using preferences directly. Given a query, the preference data $\mathcal{D}=\{(x, y_w, y_l)\}$ contains the query $x$, chosen span $y_w$, and rejected span $y_l$, and the objective of DPO is denoted as:
\begin{equation}
\begin{split}
\mathcal{L}_{DPO}  &= - \mathbb{E}_{(x,y_w,y_l) \sim \mathcal{D}}  \left[  \log \sigma \left( \beta \log\frac{\pi_{\theta}( y_w|x)}{\pi_{ref}(y_w|x))} \right. \right. \\
  & \left. \left. - \beta \log\frac{\pi_{\theta}(y_l|x)}{\pi_{ref}(y_l|x)} \right) \right]
\end{split}
\end{equation}
where $\beta$ is a parameter controlling the deviation from the base reference policy $\pi_{ref}$.

It's crucial to highlight that the construction of preference data is closely tied to business metrics. By employing a learning-to-rank approach, preference pairs such as <exposed but not clicked, clicked> and <random negative, clicked> are created, which enhances the visibility of products that are more likely to convert.

\subsection{Constrained Beam-search}

During the inference process, given a query text, the trained autoregressive language model SFT/DPO model could generate predicted identifiers in an autoregressive manner with constrained beam search, which adopts the FM-index \cite{FM_index} to identify the set of possible next tokens,  avoiding invalid identifiers without in all item title.  

More precisely, after a single decoding pass, we get a set of n-grams along with their autoregressively computed probabilities according to the model LM and then retrieve their FM-index scores via normalized index frequencies. 
The constrained beam-search score is the sum of the model score and FM-index score, formulated as 
\begin{equation}
     s(q)= f(q; b; \text{FM-index})
\end{equation}
where b is the beam size for beam search.
Subsequently, we obtained a refined probability distribution, and by employing various ranking strategies such as top@p and top@k, we generated a set of n-grams, which are the next-step inputs. This process continued until the generation phase was completed.  During this computation process, if an out-of-vocabulary (OOV) n-gram is encountered, its feature mapping (FM) score will be assigned negative infinity. As a result, it will be filtered out of the selection process. This approach falls under the retrieval-augmented paradigm(RAG), which effectively reduces the rate of hallucination, thereby enhancing the efficacy and accuracy of the inference process.
More details could refer to the original paper SEAL \cite{seal-bevilacqua2022autoregressive}.

Leveraging the above constrained beam-search, we efficiently harvest a batch of potent spans. Subsequently, we employ the FM-index for the swift identification of items that closely correspond to these segments. Importantly, the FM-index operates independently of span positioning, thus ensuring comprehensive retrieval of all relevant items, a feature that is in harmony with the objectives set forth by the task redefinition module.

\section{Experiments}


\subsection{Datasets and Metrics}
\begin{figure}
    \centering
    \includegraphics[scale=0.31]{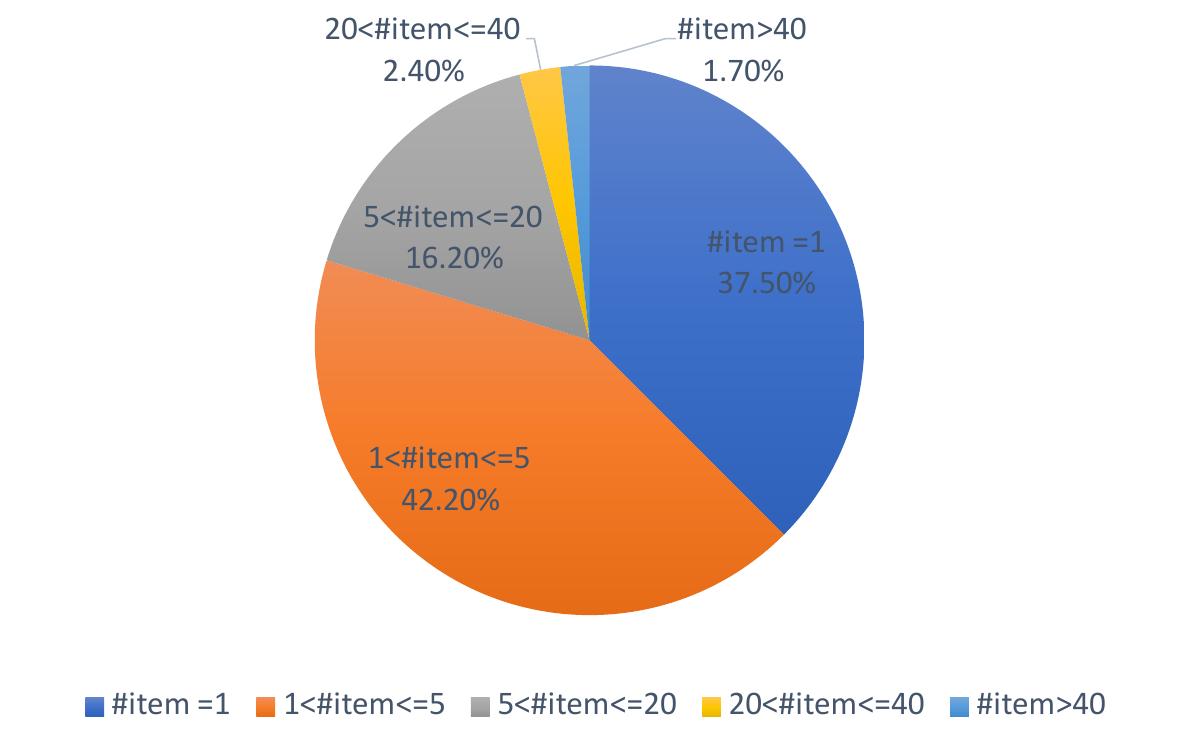}
    \caption{The distribution of percentages across different queries.}
    \label{fig:enter-label}
\end{figure}

We collect search logs of user clicks and purchases from an online E-commerce website,  where the size of the dataset is 2.8 billion. We choose the standard retrieval quality metric  Recall@K to measure the results based on the full corpus, where $K \in \{500, 1000\}$ respectively. To examine the model's performance on long-tail queries with fine granularity, we divided the original queries into five groups based on the word-level click count. As shown in Figure \ref{fig:enter-label}, queries with less than 5 clicks per account for 80\%, indicating a significant long-tail effect.
\subsection{Baselines}
In the industrial field, there are two foundational paradigms, dense retrieval and generative retrieval. Therefore, we conduct separate experimental comparisons for each paradigm.
\begin{itemize}
    \item Dense retrieval. This paradigm is the most widely used work and makes a great success.  The representative work is DSSM \cite{huang2013learning} and the variant version with a pre-trained model based on Bert~\cite{devlin2018bert}. Without loss of generality, we select \textbf{RSR}~\cite{qiu2022pre}  as the representative of the backbone of Bert, which had been deployed in the online system, severing hundreds of millions of users. 
    \item Generative retrieval.  This paradigm is an emerging and promising work. Based on different identifiers, it can be divided into two main categories, numerical-based method and lexical-based method. 
    The state-of-the-art work numerical-based is \textbf{TIGER} \cite{tiger}, which utilizes semantic codes generated by residual quantization (RQ) as identifiers. In this paper, we first use the two-tower (RSR) product of the item's embedding and then construct the semantic ID of a given item by RQ.
    The most relevant lexical-based model is \textbf{SEAL} \cite{seal-bevilacqua2022autoregressive} uses arbitrary n-grams in documents as identifiers, and retrieves documents under the constraint of a pre-built FM-indexer.
    What's more, GenR-PO is easily extensible and could be adapted in various aligning via LTR learning \cite{ppo-zhou2023enhancing,qiu2022pre,li2023learning}.
    
\end{itemize}
 
\subsection{Implementation Details}
To ensure a fair comparison among different methods, we keep the vocabulary size, the dimension of query/item, and parameters of PQ the same as \cite{mixqp}. Specifically, we set the dimension as 128, batch size as 350, n-list of IVF-PQ as 32768, nprobe as 1, and the indexing construction is used in the Faiss ANNS library\footnote{https://github.com/facebookresearch/faiss}. The default temperature $\tau$ of softmax is 1/30. The Adam optimizer is employed with an initial learning rate of 5e-5, and 6e-5 for RSR/SFT and DPO respectively. 
The default value of beam-search size is set to 100. The base model of SEAL, TIGER, and ours are all BART-large\footnote{https://huggingface.co/facebook/bart-large} \cite{lewis2019bart}. For the TIGER model, the parameter is set to RQ3x12, which consists of three residual layers, each encoding 4096 codebooks, enabling the representation of a product scale in the tens of billions.
\begin{table*}
    \centering
    \caption{The effect of different tasks for the performance.}

    \begin{tabular}{c|c|c}
    \toprule
    Task & Recall@500  & Recall@1000\\
    \midrule
       query2title  & 0.0180 &0.0240  \\
       title2query  & 0.0160 & 0.0232\\
    \midrule 
    query2multi-span  (l=10, m=2) & 0.3600 & 0.4070\\
    query2multi-span  (l=8, m=7) & \textbf{0.3680} & \textbf{0.4310}\\
    \bottomrule
    \end{tabular}
    \label{tasks}
\end{table*}

\subsection{Experiment Results}
The experimental results are shown in Table \ref{tab1}. We can conclude that the proposed framework achieves a
significant improvement over dense retrieval and generative retrieval. Specifically, 
\begin{itemize}
    \item Compared with GenR-PO + SFT,  SEAL + SFT leads to a performance decline in various metrics, showing that the straightforward query2title task is ineffective. This result is consistent with previous analysis. 
    Compared with TIGER, the lexical-based GenR-PO* makes a great improvement, indicating that it describes a better semantic match. However, the numerical pattern has a semantic gap that requires additional alignment.

    \item  Compared with RSR, GenR-PO, and variable versions perform better in terms of different long-tail queries, especially in the \#item=1.  This phenomenon indicates that generative paradigms have increased generalization capabilities compared to traditional paradigms.  
    Additionally, it is observed that the performance of the generative method varies significantly across different types of queries. For example, under head queries, the suboptimal performance of the generative method is more pronounced, which may be associated with one-to-many map learning.

    \item Through ablation studying, we can discover that the DPO  has a certain improvement in performance, especially for head queries. In addition, constrained beam-search has a significant impact on the quality of generation, indicating the important role in filtering.
\end{itemize}

\begin{figure}[]
    \centering
    \includegraphics[scale=0.36]{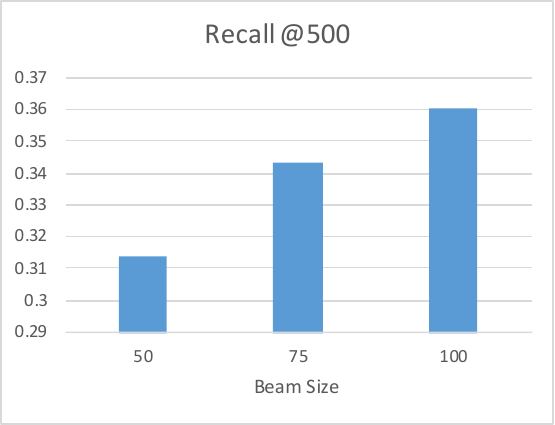}
    \includegraphics[scale=0.36]{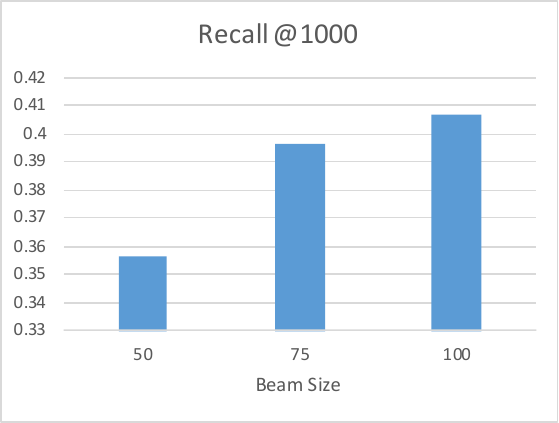}
    \caption{The performance of different beam sizes.}
    \label{beamsize}
\end{figure}


\subsection{Impact of Different Tasks}
To investigate the effect of different tasks on performance, we conduct several tasks, i.e., query2title, title2query, and query2multi-span. The results are shown in Table \ref{tasks}. We can find that the performance of the query2title and title2query tasks are extremely poor, while the query2multi-span task has significantly improved. This suggests that there is noise in the original data in the e-commerce field, which once again underscores the importance of task re-definition.

\subsection{Impact of Beam Size}
The beam size controls the quality and quantity of the generated results, impacting the model's performance. Here, we conduct additional experiments to explore the influence (parameters are l=10, m=2). The experimental results are shown in figure \ref{beamsize}. As the size increases, the effect improves, but the gradient of improvement decreases. It is also found that the larger the beam size, the greater the irrelevance of the returned results. Therefore, in practical applications, a certain compromise must be made.

\subsection{Online A/B Test}
To investigate the effectiveness of the model, we conduct online A/B testing in the online e-commerce search engine. We employ caching to store the results, which are subsequently integrated into our online recall system. Following a week of comprehensive monitoring, the outcomes are presented in Table \ref{tab:ab_test}. We can find that the performance of the proposed increases by 0.225\% in UCVR (p-value=0.0276), which demonstrates that the model provides more high-quality candidates, thereby augmenting the conversion ratio.

\begin{table}[tp]
    \centering
    \caption{Online performance of A/B tests. The improvements are averaged over a week in 2024. p-value is obtained by t-test over the online dense retrieval model.}
    \label{tab:ab_test}
    {
    \begin{tabular}{c|c|c}
    \hline
         Metric & UCVR & UV-value \\
    \hline
         Gain & +0.225\% & +0.050\%  \\
        p-value$^{\mathrm{b}}$ & 0.0276 & 0.8780  \\
    \hline
    \multicolumn{3}{l}{$^{\mathrm{b}}$  Small p-value means statistically significant.}
    \end{tabular} 
    \label{tab:abtest}
    }
\end{table}
\section{Conclusion}

This paper introduces an innovative generative retrieval framework with an optimization preference tailored for E-commerce search. The framework is crafted to adeptly train an autoregressive model in line with the target data and leverage constrained beam-search to produce the ultimate item selection. To cater to the E-commerce domain, we reconstruct the raw item titles and employ multi-span as identifiers, thereby converting the query2title task into a query2multi-span task, which simplifies the generation process. During inference, a constrained beam-search approach is utilized to pinpoint crucial spans, meaning as well as the interpretability of the retrieved items. Comprehensive testing on a real-world dataset shows that our framework markedly outperforms contemporary generative retrieval and dense retrieval in long-tail queries. The A/B test demonstrates that the model has brought about substantial conversion gains.

In future work, we aim to harness the power of large language models to bolster representation and generation capabilities for the base model and formulate an improved learning-to-rank scheme to amplify the pertinence of the generated outcomes.


\bibliography{custom}





\end{document}